\documentclass[%
 reprint,
 amsmath,amssymb,
 aps,
]{revtex4-1}

\usepackage{graphicx}
\usepackage{dcolumn}
\usepackage{bm}
\usepackage{subfigure}
\usepackage{float} 
\begin{document}

\preprint{APS/123-QED}

\title{Entanglement between light and microwave via electro-optic effect}

\author{Jinkun Liao$^{1}$}\thanks{E-mail: jkliao@uestc.edu.cn}
\author{Qizhi Cai$^{1}$}
\author{Qiang Zhou$^{1,2}$}
\address{$^{1}$School of optoelectronic science and engineering,University of Electronic Science and Technology of China, Chengdu, Sichuan, China}
\address{$^{2}$Institute of Fundamental and Frontier Sciences, University of Electronic Science and Technology of China, Chengdu, Sichuan, China}

\date{\today}

\begin{abstract}
We theoretically proposed one of the approaches achieving the quantum entanglement between light and microwave by means of electro-optic effect. Based on the established full quantum model of electro-optic interaction, the entanglement characteristics between light and microwave are studied by using the logarithmic negativity as a measure of the steady-state equilibrium operating point of the system. The investigation shows that the entanglement between light and microwave is a complicated function of multiple physical parameters, the parameters such as ambient temperature, optical detuning, microwave detuning and coupling coefficient have an important influence on the entanglement. When the system operates at narrow pulse widths and/or low repetition frequencies, it has obvious entanglement about $20$ K, which is robust to the thermal environment.   
\end{abstract}

\pacs{Valid PACS appear here}
\maketitle

\section{\label{sec:level1}Introduction}

Since the establishment of self-consistent quantum mechanics, quantum entanglement has been the focus of many famous physicists working at variety branches of physics [1-2]. Quantum entanglement not only provides a new approach for people to understand the intrinsic traits of quantum physical principles, but also becomes the source of many applications of quantum information, such as quantum computing, quantum cryptography, quantum sensing and quantum internet [3-4]. So far, a large number of quantum entanglement phenomena have been verified using microscopic and mesoscopic quantum entities, including schemes using photons, atoms, ions, and spins, etc. In recent years, protocols involving optomechanics, optoelectromechanics, microwaves and multi-mechanical oscillators have also been proposed and demonstrated in experiments [5-14]. Among these schemes, the optomechanical one is the most intensely investigated, and it has been successfully applied in gravitational wave detection [15-17]. At the same time, perspicacious physicist also noticed intuitively that the use of electro-optic effect can also achieve the quantum entanglement between light and microwave [18], but still needed to conduct more detailed and in-depth theoretical research and experimental verification.

In view of this, this paper proposed a possible scheme to obtain the entanglement between light and microwave via electro-optic effect. The electro-optic material placed in the Fabry-Perot cavity modulates the phase of the light field with the help of a microwave field, so that the output light and the microwave can generate quantum correlation or namely quantum entanglement [18]. Therefore, it has research interest and potential scientific value for quantum information science and technology. The system can obtain the entanglement between coherent light and microwave and the conversion of quantum states between light and microwave in the regime of continuous variables, and its ability to resist thermal noise interference is possibly stronger. Due to the advancement of microfabrication technology, the entangled system or device prepared by using electro-optic materials, compared with the system involving mechanical harmonic oscillators, have the advantages of small footprint, stable structure and simple preparation, and the system could be expected to be applied to quantum information processing, quantum sensing, quantum network, quantum memory and quantum interfaces, etc. [19-22]. In particular, entangled systems containing mechanical oscillators must be placed in a completely stationary state for applications, while electro-optic quantum entanglement systems can be used in fixed or moving or even accelerated environments or platforms [23-24]. 

As one of the feasible application scenarios of electro-optic entanglement, the quantum illumination scheme proposed by S. Lloyd greatly improves the signal-to-noise ratio of the target detection signal by utilizing the quantum entanglement characteristics of the detection source in a harsh thermal noise and other noise environment. The microwave quantum illumination scheme proposed by Sh. Barzanjeh et al. took S. Lloyd's protocol to a practical step [24]. Our proposed microwave and optical quantum entanglement scheme based on electro-optic effect provides an alternative for microwave quantum illumination. In addition, the scheme could enhances the correlation between quantum subsystems in a hybrid quantum network. For example, strong interaction between microwave and superconducting qubit could be realized. Photons can transmit and distribute quantum entanglement as a flying bit in the network, and can interact with quantum systems such as atoms and ions at a long distance. The electro-optic effect of quantum entanglement of microwaves and light provides a new solution for further achieving a more powerful hybrid quantum system.
    
This article is organized as follows. In Sec. II, based on the established physical model, namely the electro-optic entanglement system, the quantum Langevin equations (QLEs) describing the dynamic behavior of the system are derived by considering the influence of external damping and environmental noise. The QLEs are linearized near the stable equilibrium point of the system to obtain the corresponding Lyapunov equation. Logarithmic negativity is used as the entanglement measure for the bipartite system. In Sec. III, the numerical solution of Lyapunov equation is used to calculate the entanglement measure for investigating the influence of various physical parameters on entanglement. Typically, the effects of ambient temperature, optical detuning, microwave detuning, optical power and microwave power on the entanglement are studied. The physical mechanism behind the numerical result is simply analysed in sec. IV, while Sec. V is for conclusion.
\section{\label{sec:level1} EO entanglement system and quantum Langevin equations}
The structure of the electro-optic entanglement system is shown in Fig. 1. The incident laser beam enters the optical cavity and passes through the electro-optic material in the optical cavity. Due to the electro-optic effect, the microwave signal modulates the refractive index of the material to modulate the phase of the laser. The phase-modulated laser is reflected from the optical cavity and has quantum entanglement with the output microwave from the electro-optic material. The optical resonant cavity generally adopts a Fabry-Perot cavity, wherein one cavity mirror is partially reflective, and the other cavity mirror is totally reflective. The microwave resonant cavity uses a superconducting microwave resonant circuit, wherein the transmission line adopts a microstrip line or a coplanar line, depending on the specific electro-optic materials such as inorganic electro-optic crystal or organic polymer, respectively. As depicted in Fig. 1, the electro-optic entanglement system consists of an optical cavity, a microwave resonant circuit and an electro-optic material, wherein the resonant frequency of the microwave oscillating circuit is $\omega _{m}$ and the resonant frequency of the optical cavity is $\omega _{o}$. The microwave resonant circuit is equivalent to a superconducting LC oscillating circuit. The alternating electric field applied to the electro-optic material changes its refractive index to make electro-optic interaction. The intensity of this electro-optic interaction can be described by the coupling coefficient as follows [18]
\begin{eqnarray}
g = \frac{{{\omega _{do}}{n^3}rl}}{{c\tau d}}{\left( {\frac{{\hbar {\omega _{dm}}}}{{2C}}} \right)^{1/2}}
\end{eqnarray}
where $n$ is the refractive index of the electro-optic material, $r$ is the electro-optic coefficient, $l$ is the electro-optical interaction length, $d$ is the electrode spacing, $\tau$ is the round-trip time of the light wave in the F-P cavity, and $C$ is the equivalent capacitance, respectively and $c$ is the speed of light in free space; $\omega _{do}=\omega _{o}-\Delta _{o}$ and $\omega _{dm}=\omega _{m}-\Delta _{m}$ are the driving frequencies of light waves and microwaves, respectively. $\Delta _{o}$ and $\Delta _{m}$ are the detuning of light and microwave,respectively.
\begin{figure}
	\centering
	\includegraphics[width=1\linewidth]{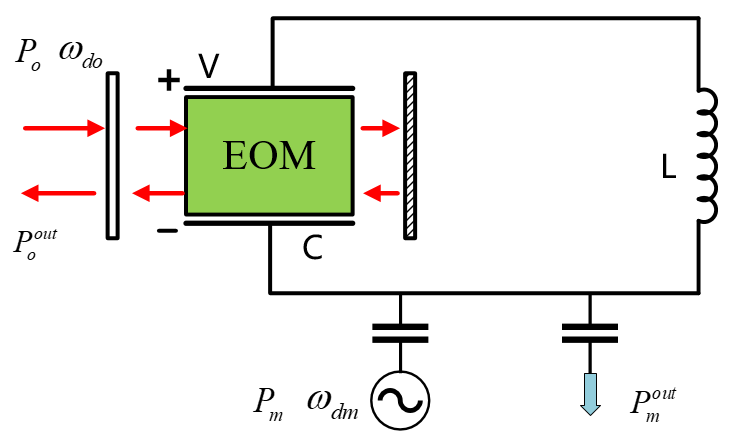}
	\caption{Schematic diagram of the structure of electro-optic entanglement system [18].}
	\label{fig:Fig1.png}
\end{figure}

As shown in Fig. 1, under the adiabatic approximation, i.e. ${\omega _m} <  < c/2nL$ , the Casimir effect, retardation, and Doppler effect are ignored [2]. The Hamiltonian of the electro-optical entanglement system is [5-7]
\begin{widetext}
    \begin{eqnarray}
	H = \hbar {\omega _o}a_o^\dag {a_o} + \hbar {\omega _m}a_m^\dag {a_m} - \hbar g(a_m^\dag  + {a_m})a_o^\dag {a_o} + i\hbar {E_o}(a_o^\dag {e^{ - i{\omega _{do}}t}} - {a_o}{e^{i{\omega _{do}}t}}) - i\hbar {E_m}({e^{i{\omega _{dm}}t}} - {e^{ - i{\omega _{dm}}t}})({a_m} + a_m^\dag )
    \end{eqnarray}
\end{widetext}
where $a_{o}$,$a_{m}$,$a_{o}^{\dagger }$,$a_{m}^{\dagger }$ are the annihilation and creation operators corresponding to the optical field and the microwave field,respectively.
\begin{eqnarray}
 {E_o} = \sqrt {2{P_o}{\gamma _o}/\hbar {\omega _{do}}},
 {E_m} = \sqrt {2{P_m}{\gamma _m}/\hbar {\omega _{dm}}}
\end{eqnarray}
where $E_{o}$,$E_{m}$ are respectively the equivalent driving strength related to optical field and microwave field intensity. $P_{o}$,$P_{m}$ are the driving power of light and microwave, $\gamma_{o}$,$\gamma_{m}$ are the damping rates of light and microwave,respectively. Regarding the interaction picture of ${H_o} = \hbar {\omega _{do}}a_o^\dag {a_o} + \hbar {\omega _{dm}}a_m^\dag {a_m}$, and considering the rotating wave approximation, ignore the fast oscillation terms $\pm 2{\omega _{do}}$ and $\pm 2{\omega _{dm}}$ ,we have 
\begin{eqnarray}
\begin{aligned}
   H = &\hbar {\Delta _o}a_o^\dag {a_o} + \hbar {\Delta _m}a_m^\dag {a_m} - \hbar g(a_m^\dag  + {a_m})a_o^\dag {a_o} \\&+ i\hbar {E_o}(a_o^\dag  - {a_o}) - i\hbar {E_m}({a_m} - a_m^\dag )   
\end{aligned}
\end{eqnarray}
Thus, we can write the respective Heisenberg equations that operators $a_{o}$,$a_{m}$ satisfy.

The whole quantum system is inevitably affected by environment, so the damping and noise terms are added to the Heisenberg equations of motion of ${a_o},{a_m}$ phenomenologically. It is not difficult to introduce the nonlinear quantum Langevin equations (QLEs) to describe the quantum dynamic behavior of light and microwave fields.
\begin{eqnarray}
\begin{aligned}
{\dot a_o} =  &- i{\Delta _o}{a_o} + ig({a_m} + a_m^\dag ){a_o} \\&+ {E_o} - {\gamma _o}{a_o} + \sqrt {2{\kappa _o}} {a_{o,in}}
\end{aligned}
\end{eqnarray}
\begin{eqnarray}
\begin{aligned}
{\dot a_m} =  &- i{\Delta _m}{a_m} + iga_o^\dag {a_o} \\&+ {E_m} - {\gamma _m}{a_m} + \sqrt {2{\kappa _m}} {a_{m,in}}
\end{aligned}
\end{eqnarray}
where $a_{o,in}$ and $a_{m,in}$ are the input noise of light and microwave, respectively, and can theoretically be regarded as a Gaussian process with zero mean, satisfying the following correlation [25]
\begin{eqnarray}
\left\langle {{a_{o,in}}(t)a_{o,in}^\dag (t')} \right\rangle  = [N({\omega _o}) + 1]\delta (t - t')
\end{eqnarray}
\begin{eqnarray}
\left\langle {a_{o,in}^\dag (t){a_{o,in}}(t')} \right\rangle  = N({\omega _o})\delta (t - t')
\end{eqnarray}
\begin{eqnarray}
\left\langle {{a_{m,in}}(t)a_{m,in}^\dag (t')} \right\rangle  = [N({\omega _m}) + 1]\delta (t - t')
\end{eqnarray}
\begin{eqnarray}
\left\langle {a_{m,in}^\dag (t){a_{m,in}}(t')} \right\rangle  = N({\omega _m})\delta (t - t')
\end{eqnarray}
where $N(\omega _{o})=[e^{\hbar\omega _{o}/k_{B}T}-1]^{-1}$ and $N(\omega _{m})=[e^{\hbar\omega _{m}/k_{B}T}-1]^{-1}$ are the light wave and microwave excitation numbers, respectively, and $k_{B}$ is the Boltzmann constant.

Let ${\dot a_o}{\rm{ = }}0,{\dot a_m}{\rm{ = }}0$,${a_o} \to {\alpha _o},{a_m} \to {\alpha _m}$ and ignore the input noise in Eq.(4) and Eq.(5), we can get the constraint relationship that the stable equilibrium operating point satisfies
\begin{eqnarray}
 - i{\Delta _o}{\alpha _o} + ig({\alpha _m} + \alpha _m^\dag ){\alpha _o} + {E_o} - {\gamma _o}{\alpha _o} = 0
\end{eqnarray}
\begin{eqnarray}
- i{\Delta _m}{\alpha _m} + ig\alpha _o^\dag {\alpha _o} + {E_m} - {\gamma _m}{\alpha _m} = 0
\end{eqnarray}
Solving Eq.(11) and Eq.(12) can obtain a stable equilibrium point of the electro-optic entanglement system $({\alpha _o},{\alpha _m})$.

To facilitate mathematical processing, we linearize equations Eq.(5) and Eq.(6) near the stable equilibrium point of the system. Therefore, let $a_{o}=\alpha _{o}+\delta a_{o}$ and $a_{m}=\alpha _{m}+\delta a_{m}$, and substituting them into equations Eq.(5) and Eq.(6), where $\delta a_{o}$,$\delta a_{m}$ represent the fluctuations of the light field and the microwave field. When the light and the microwave driving signals are strong enough, there are $\left | \alpha _{o} \right |\gg 1$,$\left | \alpha _{m} \right |\gg 1$, which can linearize the dynamic equation near the steady state safely, that is, ignoring the term of the second or higher order fluctuations, the linear quantum Langevin equations of the light and the microwave quantum fluctuations are bellow as
\begin{eqnarray}
\begin{aligned} 
\delta {\dot a_o} =& (2i{\alpha _m}g - i{\Delta _o} - {\gamma _o})\delta {a_o} \\&+ ig{\alpha _o}(\delta {a_m} + \delta a_m^\dag ) + \sqrt {2{\kappa _o}} \delta {a_{o,in}}
\end{aligned} 
\end{eqnarray}
\begin{eqnarray}
\begin{aligned} 
\delta {\dot a_m} =&  - (i{\Delta _m} + {\gamma _m})\delta {a_m} \\&+ ig{\alpha _o}(\delta {a_o} + \delta a_o^\dag ) + \sqrt {2{\kappa _m}} \delta {a_{m,in}}
\end{aligned}
\end{eqnarray}

Introducing the orthogonal operators of the fluctuation of the light field and the microwave field,the orthogonal operators of the light field fluctuation are
\begin{subequations}  \label{eq:1}
\begin{align}  
\delta X_{o}=(\delta a_{o}+\delta a_{o}^{\dagger })/\sqrt{2}\label{eq:1A} \\
\delta Y_{o}=(\delta a_{o}-\delta a_{o}^{\dagger })/i\sqrt{2}\label{eq:1B}
\end{align}
\end{subequations}

The orthogonal operators of the microwave field fluctuation are
\begin{subequations}  \label{eq:2}
\begin{align}  
\delta X_{m}=(\delta a_{m}+\delta a_{m}^{\dagger })/\sqrt{2}\label{eq:2A}\\
\delta Y_{m}=(\delta a_{m}-\delta a_{o}^{\dagger })/i\sqrt{2}\label{eq:2B}
\end{align}
\end{subequations}

Similarly, the corresponding optical and microwave fields’ input noise fluctuation operators are
\begin{subequations}  \label{eq:2}
\begin{align}  
\delta A_{o}^{in}=(\delta a_{o,in}+\delta a_{o,in}^{\dagger })/\sqrt{2}\label{eq:2A}\\
\delta B_{o}^{in}=(\delta a_{o,in}-\delta a_{o,in}^{\dagger })/i\sqrt{2}\label{eq:2B}
\end{align}
\end{subequations}
and
\begin{subequations}  \label{eq:2}
\begin{align}  
\delta A_{m}^{in}=(\delta a_{m,in}+\delta a_{m,in}^{\dagger })/\sqrt{2}\label{eq:2C}\\
\delta B_{m}^{in}=(\delta a_{m,in}-\delta a_{m,in}^{\dagger })/i\sqrt{2}\label{eq:2D}
\end{align}
\end{subequations}

After linearization, QLEs for fluctuations can be written as
\begin{subequations}  \label{eq:2}
\begin{align}  
&\delta \dot{X}_{o}=-\gamma _{o}\delta X_{o}+(\Delta _{o}-2g\alpha _{m})\delta Y_{o}+\sqrt{2\kappa _{o}}\delta A_{o}^{in}\label{eq:2A}
\\&\delta \dot{Y}_{o}=(2g\alpha _{m}-\Delta _{o})\delta X_{o}-\gamma _{o}\delta Y_{o}+2g\alpha _{o}\delta X_{m}+\sqrt{2\kappa _{o}}\delta B_{o}^{in}\label{eq:2B}
\\&\delta \dot{X}_{m}=\Delta_{m}\delta Y_{m}-\gamma_{m}\delta X_{m}+\sqrt{2\kappa _{m}}\delta A_{m}^{in}\label{eq:2C}
\\&\delta \dot{Y}_{m}=2g\alpha _{o}\delta X_{o}-\Delta_{m}\delta X_{m}-\gamma _{m}\delta Y_{m}+\sqrt{2\kappa _{m}}\delta B_{m}^{in}\label{eq:2D}
\end{align}
\end{subequations}
The above Eq.(19) can be written in the following matrix form simply
\begin{eqnarray}
\dot{u}(t)=Au(t)+n(t)
\end{eqnarray}
where \[u^{T}(t)=\begin{pmatrix}
\delta X_{o},\delta Y_{o},\delta X_{m},\delta Y_{m}
\end{pmatrix}\]\[n^{T}(t)=\begin{pmatrix}
\sqrt{2\kappa _{o}}\delta A_{o}^{in},\sqrt{2\kappa _{o}}\delta B_{o}^{in},\sqrt{2\kappa _{m}}\delta A_{m}^{in},\sqrt{2\kappa _{m}}\delta B_{m}^{in}
\end{pmatrix}\]
\[A=\begin{pmatrix}
-\gamma_{o} & \Delta _{o}-2g\alpha _{m} & 0 & 0 \\ 
2g\alpha _{m}-\Delta _{o} & -\gamma_{o} & 2g\alpha _{o} & 0\\ 
0 & 0 & -\gamma_{m} & \Delta _{m}\\ 
2g\alpha _{o} & 0 & -\Delta _{m} & -\gamma_{m}
\end{pmatrix}\]

The solution of equation Eq.(20) is
\begin{eqnarray}
u(t)=M(t)u(0)+\int _{0}^{t}ds M(s)n(t-s)
\end{eqnarray}
where $M(s)=exp(As)$. The steady state of the system can be characterized by the correlation matrix of the elements $V_{ij}=\left \langle u_{i}(\infty)u_{j}(\infty)+u_{j}(\infty)u_{i}(\infty) \right \rangle/2$, which can be calculated as $V=\int _{0}^{\infty}ds M(s)DM^{T}(s)$, where $D=Diag[\gamma _{o},\gamma _{o},\gamma _{m}(2\bar{n}_{b}+1),\gamma _{m}(2\bar{n}_{b}+1)]$, $\bar{n}_{b}$ is the mean thermal excited number of microwave field.

When the stability condition is satisfied, Eq.(20) is equivalent to the Lyapunov equation [9-10] as bellow
\begin{eqnarray}
AV+VA^{T}=-D
\end{eqnarray}
The four conditions that need to be met by the system stability are listed in the appendix by the Routh-Hurwitz criterion [26,27].

For the case of continuous variables, entanglement can be measured by defining logarithmic negativity [28]
\begin{eqnarray}
E_{N}=max[0,-ln2\eta ^{-}]
\end{eqnarray}
where
\[\eta ^{-}\equiv 2^{-1/2}\left \{ \sum (V)-[\sum (V)^{2}-4detV]^{-1/2} \right \}^{-1/2}\]
\[V=\begin{pmatrix}
V_{11} & V_{12}\\ 
V_{12}^{T} & V_{22}
\end{pmatrix}\]
and $\sum (V)\equiv detV_{11}+detV_{22}-2detV_{12}$. In the above system Eq.(20), if the real parts of all the eigenvalues of the matrix $A$ are negative, the whole entanglement system is stable and tends to be steady state. In the numerical calculation of the next section, the Routh-Hurwitz criterion is satisfied, that is, it is a assumed that the parameters are valued within the scope in which the stability condition is fullfilled.

\section{\label{sec:level1} Entanglement analysis at steady state}
In this paper, the commonly used laser wavelength and microwave frequency are selected to achieve electro-optic entanglement as large as possible for the system. The laser wavelength is $\lambda =1064$ nm, the microwave frequency is set as $f =9$ GHz, and the electro-optical crystal material is taken as an example of lithium niobate. Of course, other inorganic crystal materials or even organic polymer electro-optic materials can be used. At the wavelength of $1064$ nm, the electro-optic coefficient of lithium niobate is  $r=32pm/V$, and the refractive index $n=2.232$, F-P cavity length $L = 2.1$ mm, electro-optical crystal length $l = 2$ mm, thickness of electro-optic crystal $d = 50$ $\mu m$, equivalent capacitance of microwave resonant circuit $ C_{o} = 1$ pF, the characteristic parameters and other parameters of lithium niobate are respectively indicated in the title map of each sheet. In this paper, a large number of numerical calculations have been carried out to study the dependence of logarithmic negativity (i.e. entanglement) of electro-optical entanglement system on optical wave detuning, light wave power, microwave detuning, microwave power and ambient temperature. The results are shown in Figures 2-8:
\begin{figure}[htbp]
	\centering 
	\subfigure[Optical damping rates $\gamma_{o} = 0.015,0.02,0.03 \omega_{m}$ are different,and microwave damping rates are $\gamma_{m} =0.0005 \omega_{m}$.]{ 
	\includegraphics[width=3.4in]{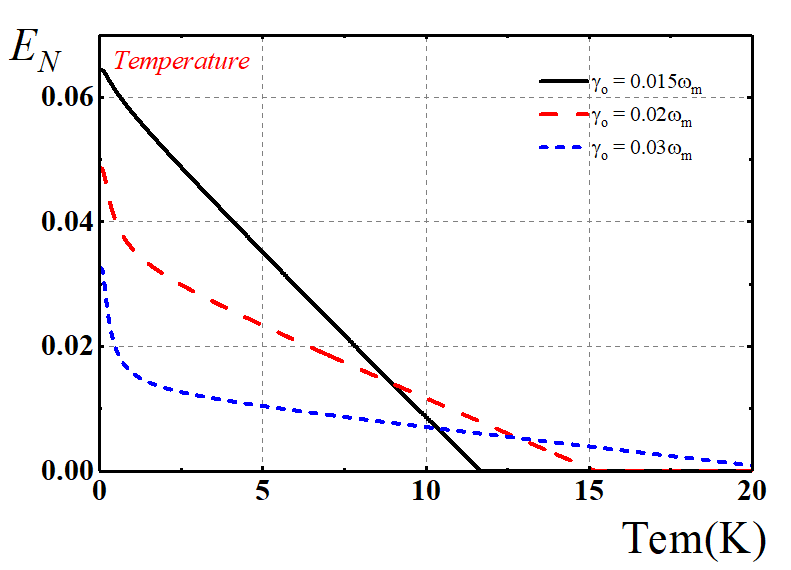} }
	\subfigure[Microwave damping rates $\gamma_{m} = 0.0005,0.001,0.0015 \omega_{m}$ are different,and optical damping rates are $\gamma_{o} =0.02 \omega_{m}$ .]{ 
	\includegraphics[width=3.4in]{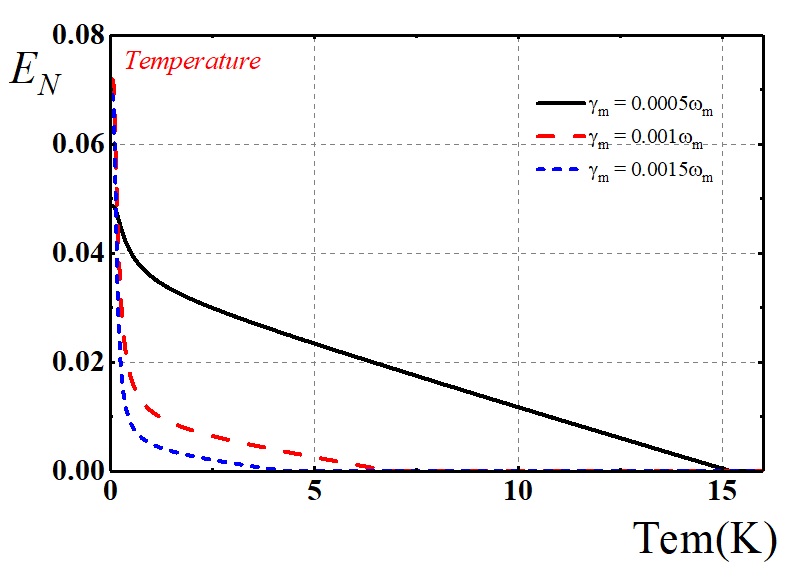} }
	\caption{Relationship between entanglement and ambient temperature. The common simulated parameters: optical wave resonator resonance wavelength $ \lambda = 1064$ nm, microwave resonator resonance frequency $f = 9$ GHz, lithium niobate refractive index $n = 2.232$, driving light power $P_{o} = 200$ mW, driving microwave power $P_{m} = 200$ mW, optical detuning coefficient $\Delta_{o}/\omega_{m} = 0.002$, microwave detuning coefficient $\Delta_{m}/\omega_{m} = 0.055$.} 
	\label{fig:Fig2.png}
\end{figure}
\begin{figure}[htbp]
	\centering 
	\subfigure[Optical damping rates $\gamma_{o} = 0.005,0.008,0.01 \omega_{m}$ are different,and microwave damping rates are $\gamma_{m} =0.002 \omega_{m}$ .]{ 
	\includegraphics[width=3.4in]{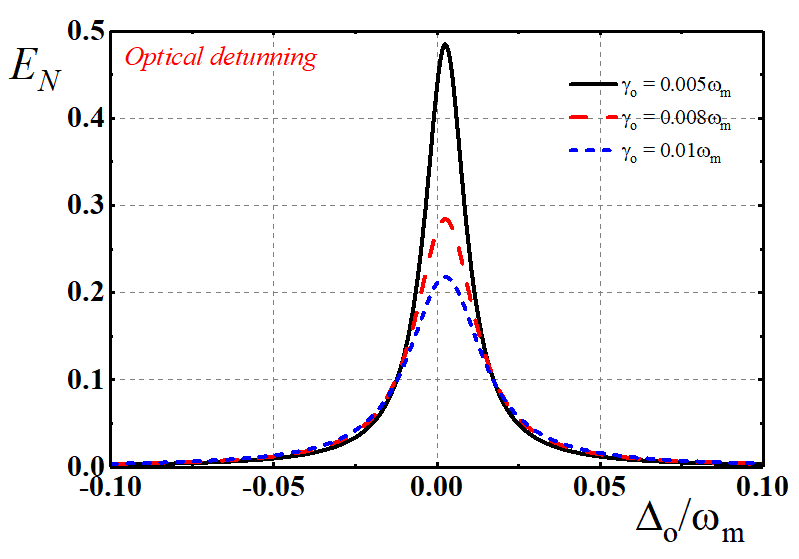} }
	\subfigure[Microwave damping rates $\gamma_{m} = 0.002,0.005,0.01 \omega_{m}$ are different,and optical damping rate are $\gamma_{o} =0.005 \omega_{m}$ .]{ 
	\includegraphics[width=3.4in]{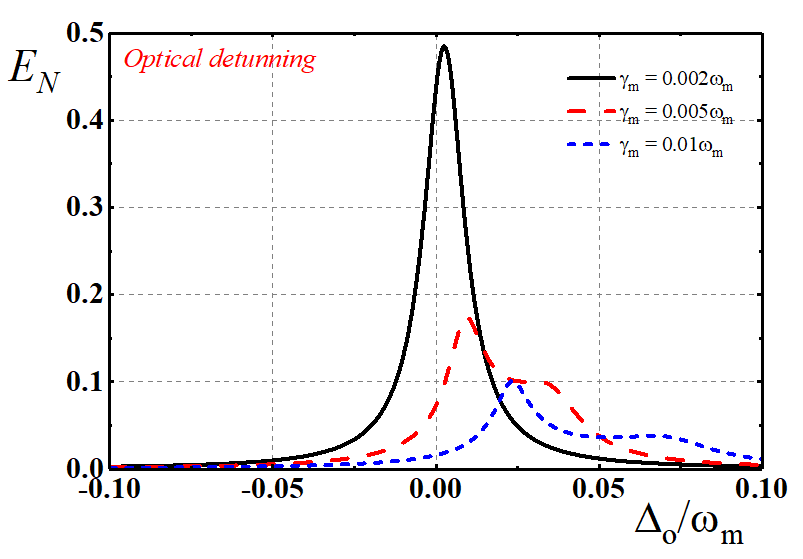} }
	\caption{Relationship between entanglement and optical detuning coefficient. The common simulated parameters: optical wave resonator resonance wavelength $ \lambda = 1064$ nm, microwave resonator resonance frequency $f = 9$ GHz, lithium niobate refractive index $n = 2.232$,temperature $T = 15$ mK, driving light power $P_{o} = 30$ mW, driving microwave power $P_{m} = 30$ mW, microwave detuning coefficient $\Delta_{m}/\omega_{m} = 0.055$.}
	\label{fig:Fig3.png}
\end{figure}
\begin{figure}[htbp]
	\centering 
	\subfigure[Optical damping rates $\gamma_{o} = 0.005,0.008,0.01 \omega_{m}$ are different,and microwave damping rate are $\gamma_{m} =0.002 \omega_{m}$ .]{ 
		\includegraphics[width=3.4in]{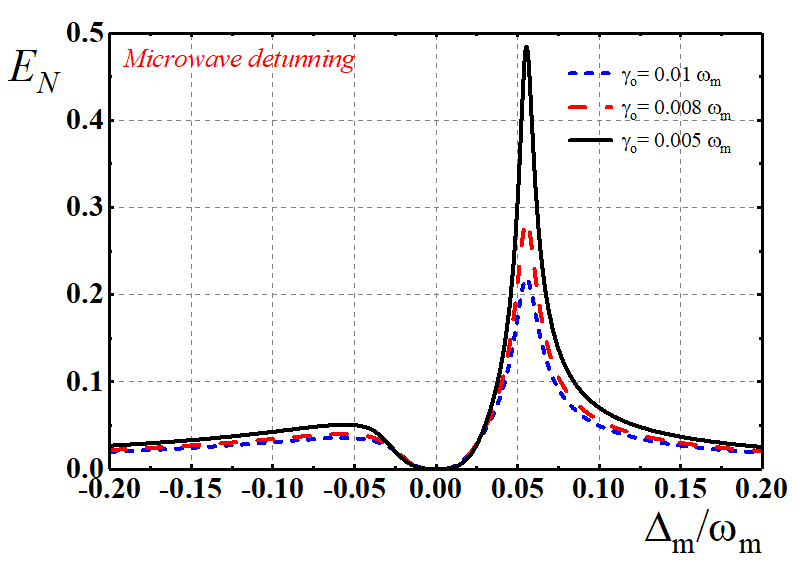} }
	\subfigure[Microwave damping rates $\gamma_{m} = 0.002,0.005,0.01 \omega_{m}$ are different,and optical damping rate are $\gamma_{o} =0.005 \omega_{m}$ .]{ 
		\includegraphics[width=3.4in]{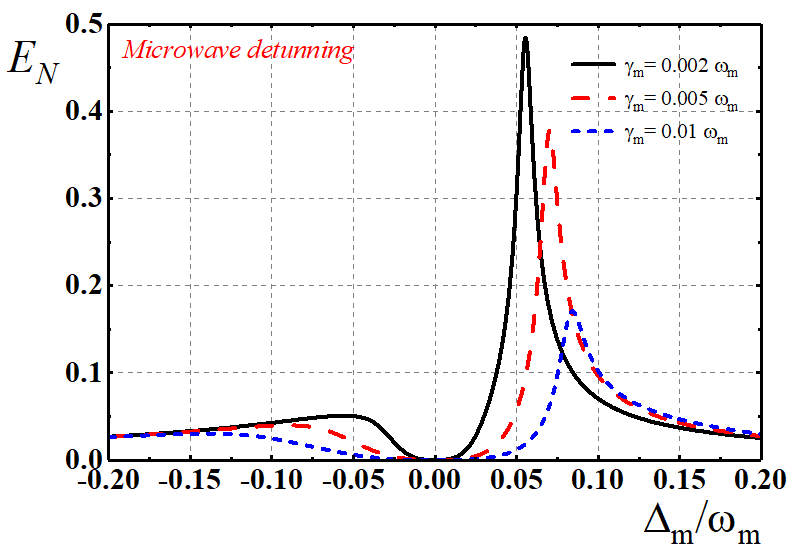} }
	\caption{Relationship between entanglement and microwave detuning coefficient. The common simulated parameters: optical wave resonator resonance wavelength $ \lambda = 1064$ nm, microwave resonator resonance frequency $f = 9$ GHz, lithium niobate refractive index $n = 2.232$,temperature $T = 15$ mK, driving light power $P_{o} = 30$ mW, driving microwave power $P_{m} = 30$ mW, optical detuning coefficient $\Delta_{o}/\omega_{m} = 0.002$.}
	\label{fig:Fig4.png}
\end{figure}
\begin{figure}
	\includegraphics[width=1\linewidth]{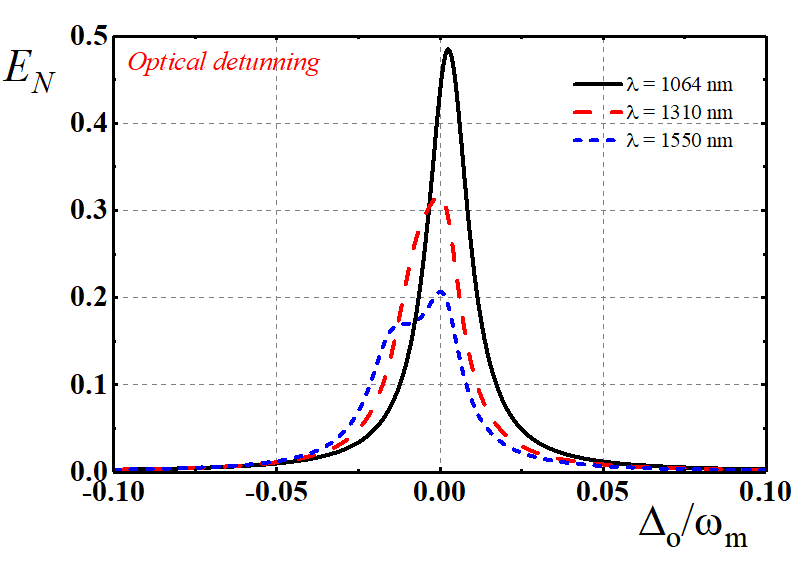}
	\caption{Relationship between entanglement and optical detuning coefficient in different driving optical wavelengths. The common simulated parameters: optical wave resonator resonance wavelength $ \lambda = 1064, 1310, 1550$ nm, microwave resonator resonance frequency $f = 9$ GHz, lithium niobate refractive index $n = 2.232,2.220,2.211$, temperature $T = 15$ mK, optical damping rate $\gamma_{o} = 0.005\omega_{m}$, microwave damping rate $\gamma_{m} =0.002 \omega_{m}$, driving light power $P_{o} = 30$ mW, driving microwave power $P_{m} = 30$ mW, microwave detuning coefficient $\Delta_{m}/\omega_{m} = 0.055$.}
	\label{fig:Fig5.png}
\end{figure}

It can be seen from Fig.2(a) that under the premise of other parameters, the entanglement of the electro-optical entanglement system decreases with the increase of the ambient temperature. When the quality factor of the optical cavity is large, the entanglement is generally large, and when the quality factor of the optical cavity is small, the entanglement is generally small but slows down with temperature slowly. As shown in Fig.2(b), the situation of the microwave cavity is different. When the quality factor of the microwave cavity is large, the entanglement decreases slowly with temperature, but when the quality factor of the microwave cavity is small, the entanglement appears to be larger about 0 K. The value is only decaying too fast with temperature. In general, the entanglement of the electro-optical entanglement system is still decreasing as the ambient temperature increases while other parameters are fixed. These numerical results show that the high-quality optical cavity and microwave cavity generally make the electro-optic entanglement more resistant to thermal noise environment, but the quality factor of the optical cavity and the microwave cavity need to be optimized or tradeoff to make the entanglement system have greater entanglement and resistant ability to temperature.
\begin{figure}
	\includegraphics[width=1\linewidth]{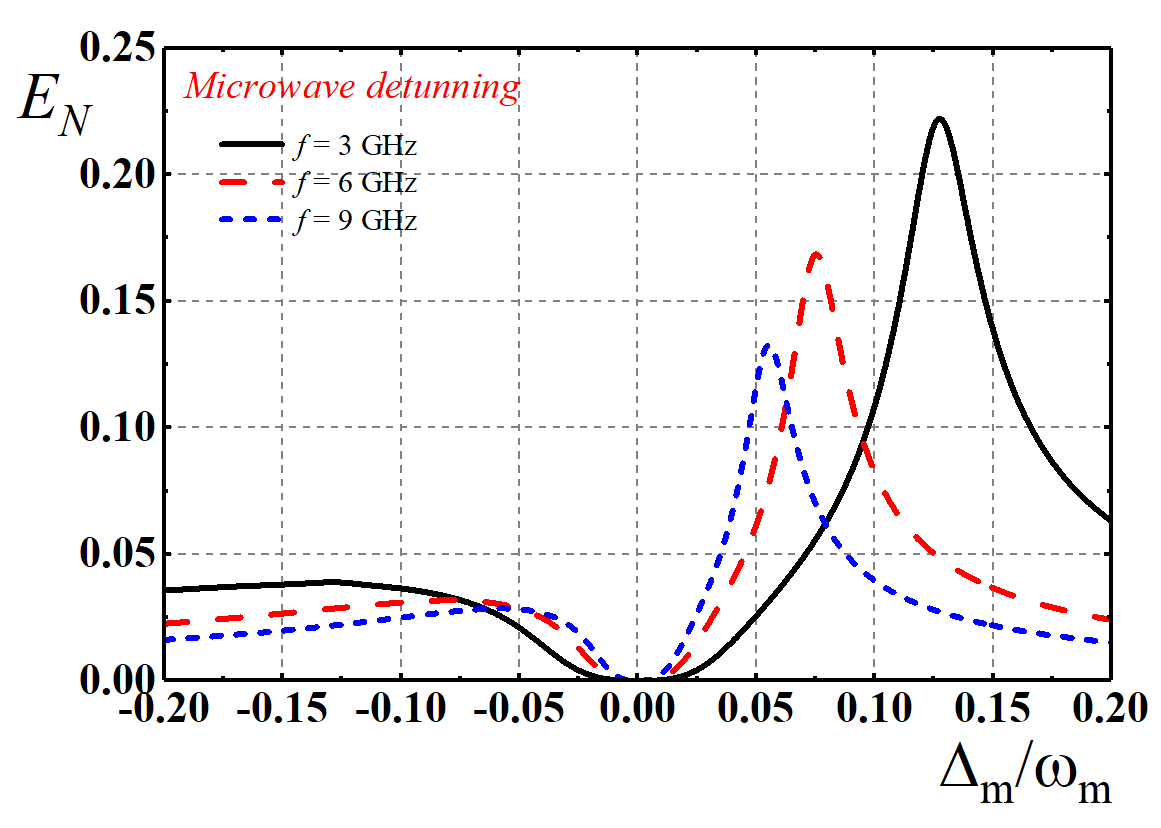}
	\caption{Relationship between entanglement and microwave detuning coefficient in different driving microwave frequencies. The common simulated parameters: optical wave resonator resonance wavelength $ \lambda = 1064$ nm, microwave resonator resonance frequency $f = 3, 6, 9$ GHz, lithium niobate refractive index $n = 2.232$, temperature $T = 15$ mK, optical damping rate $\gamma_{o} = 0.015\omega_{m}$, microwave damping rate $\gamma_{m} =0.002 \omega_{m}$, driving light power $P_{o} = 30$ mW, driving microwave power $P_{m} = 30$ mW, optical detuning coefficient $\Delta_{o}/\omega_{m} = 0.002$.}
	\label{fig:Fig6.png}
\end{figure}

In Fig.3(a), under fixed microwave relaxation coefficient, the entanglement with optical detuning is larger when the relaxation coefficient of the optical cavity is smaller, its maximum value appears near the resonant frequency of the optical cavity. As shown in Fig.3(b), under fixed optical relaxation coefficient, when the quality factor of the microwave cavity is large, the entanglement is large, the maximum value of the entanglement appears “red shift” with the decrease of the microwave quality factor. In general, when other parameters are fixed and the quality factor of the optical cavity and the microwave cavity is large, the entanglement is generally large, but the dependence of the entanglement on the relaxation coefficient of the optical cavity and the microwave cavity is different.

As shown in Fig.4(a), when the microwave relaxation coefficient is fixed, the quality factor of the optical cavity is increased to obtain greater entanglement, and the maximum value of the entanglement appears in the red sideband of the optical resonance frequency. It can be seen from Fig.4(b) that when the optical relaxation coefficient is fixed, the quality factor of the microwave cavity can be increased to achieve larger entanglement, and the maximum value of the entanglement is further “red shifted” with the decrease of the microwave quality factor. Similar to the case of entanglement with optical detuning, when other parameters are fixed and the quality factor of the optical cavity and the microwave cavity is large, the entanglement of the electro-optic system is generally large.

As shown in Fig.5, under other conditions fixed, the entanglement of the electro-optical entanglement system changes with the change of the wavelength of the driving light for the change of the optical detuning, and the entanglement decreases with the increase of the wavelength of the light, and the maximum position and trend of the values are basically unchanged. Compared with Fig.5, Fig.6 shows that the entanglement tends to zero when the microwave is detuned to zero and the entanglement is much stronger than the blue sideband for the red sideband. Futhermore, we can see that the lower the microwave frequency, the stronger the entanglement.

It can also be seen from Fig.7 that the entanglement increases slowly as the light wave drive power increases under certain other parameters, and the entanglement is relatively larger when the wavelength is shorter. As shown in Fig.8, the entanglement at 9 GHz monotonously decreases with the increase of microwave power, and the entanglement at 6 GHz and 3 GHz is maximized at some special values of microwave power. As the microwave power continues to increase, the entanglement is still monotonously decreasing.
\begin{figure}
	\centering
	\includegraphics[width=1\linewidth]{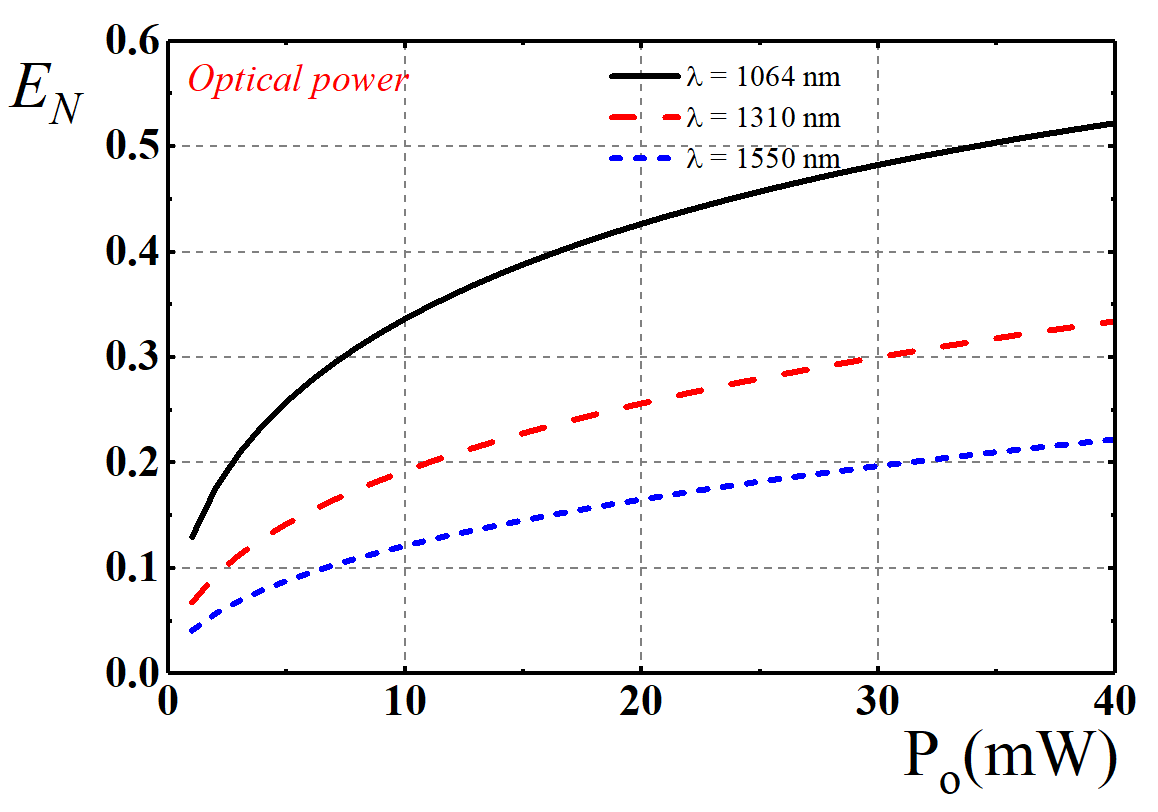}
	\caption{Optical power and entanglement. The simulated parameters: optical wave resonator resonance wavelength $ \lambda = 1064, 1310, 1550$ nm, microwave resonator resonance frequency $f = 9$ GHz, lithium niobate refractive index $n = 2.232,2.220,2.211$, temperature $T = 15$ mK, optical damping rate $\gamma_{o} = 0.005 \omega_{m}$, microwave damping rate $\gamma_{m} =0.002 \omega_{m}$, driving microwave power $P_{m} = 30$ mW,optical detuning coefficient $\Delta_{o}/\omega_{m} = 0.002$, microwave detuning coefficient $\Delta_{m}/\omega_{m} = 0.055$.}
	\label{fig:Fig7.png}
\end{figure}
\begin{figure}
	\centering
	\includegraphics[width=1\linewidth]{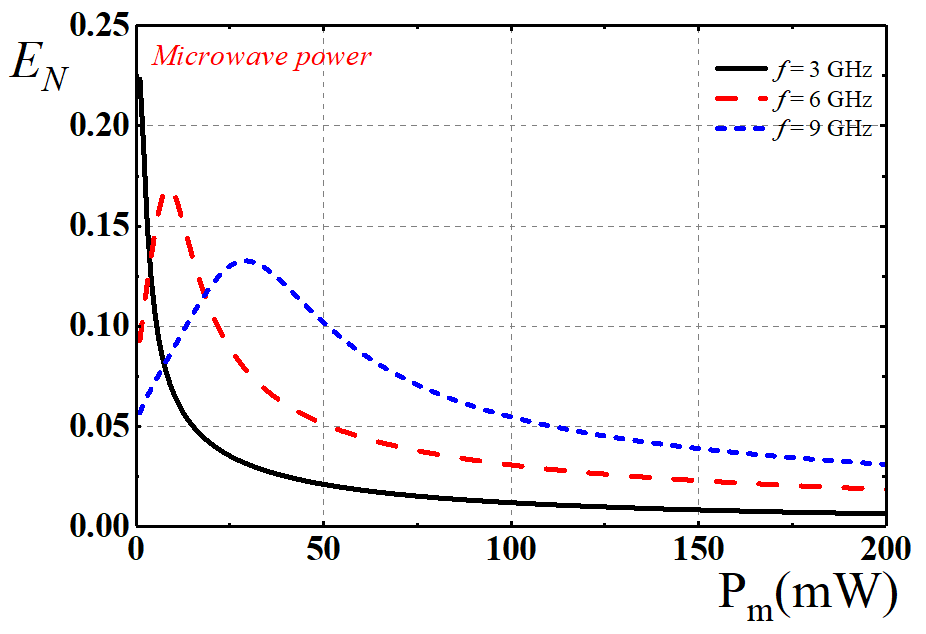}
	\caption{Microwave power and entanglement.The simulated parameters: optical wave resonator resonance wavelength $ \lambda = 1064$ nm, microwave resonator resonance frequency $f = 3, 6, 9$ GHz, lithium niobate refractive index $n = 2.232$, temperature $T = 15$ mK, optical damping rate $\gamma_{o} = 0.015 \omega_{m}$, microwave damping rate $\gamma_{m} =0.002 \omega_{m}$, driving light power $P_{o} = 30$ mW, optical detuning coefficient $\Delta_{o}/\omega_{m} = 0.002$,microwave detuning coefficient $\Delta_{m}/\omega_{m} = 0.055$.}
	\label{fig:Fig8.png}
\end{figure}

In addition, the coupling strength of the system also affects the entanglement, and the coupling strength has many influencing factors. The coupling strength can be adjusted by multiple parameter regime to achieve a suitable value, and an adjustable parameter is provided for entanglement at a higher temperature.

\section{\label{sec:level1}Result analysis}
Since the electro-optical entanglement system works in the ultra-low temperature state, the input noise can be ignored in the dynamic analysis, which we can equivalently ignore the input noise operator of the light wave and the microwave in Eq.(13) and Eq.(14), and obtains the following dynamic equation
\begin{eqnarray}
\begin{aligned}
\delta {\dot a_o} = (2i{\alpha _m}g - i{\Delta _o} - {\gamma _o})\delta {a_o} + ig{\alpha _o}(\delta {a_m} + \delta a_m^\dag )
\end{aligned}
\end{eqnarray}
\begin{eqnarray}
\begin{aligned}
\delta {\dot a_m} =  - (i{\Delta _m} + {\gamma _m})\delta {a_m} + ig{\alpha _o}(\delta a_o^\dag  + \delta {a_o})
\end{aligned}
\end{eqnarray}
Since $\delta {a_o}$ and $\delta {a_m}$ are fluctuations of light waves and microwaves, respectively, when light waves and microwaves have not been input, that is $t = 0$, they have approximately
\begin{eqnarray}
\begin{aligned}
\delta {a_o} \to 0,\delta {a_m} \to 0
\end{aligned}
\end{eqnarray}
Under the above initial conditions, the fluctuations of light and microwave in Eq.(24) and Eq.(25) can be approximated as follows
\begin{subequations}  \label{eq:2}
\begin{align}  
&\delta {a_m} \propto {e^{ - ({\gamma _m} + i{\Delta _m})t}}\label{eq:2A}
\\&\delta a_m^\dag  \propto {e^{ - ({\gamma _m} - i{\Delta _m})t}}\label{eq:2B}
\\&\delta {a_o} \propto {e^{ - {\gamma _o}t}}{e^{i(2g{\alpha _m} - {\Delta _o})t}}\label{eq:2C}
\\&\delta a_o^\dag  \propto {e^{ - {\gamma _o}t}}{e^{ - i(2{\alpha _m}g - {\Delta _o})t}}\label{eq:2D}
\end{align}
\end{subequations}
It is not difficult to get
\begin{equation}
\begin{split}
\delta {a_o}(t)& \propto ig{\alpha _o}{e^{[ - {\gamma _o} - i({\Delta _o} + 2g{\alpha _m})]t}}\cdot\\&\int\limits_0^t {ds{e^{({\gamma _o} - {\gamma _m})s}}[{e^{i{\Delta _m}s}} + {e^{ - i{\Delta _m}s}}]{e^{i({\Delta _o} - 2g{\alpha _m})s}}}
\end{split}
\end{equation}
\begin{equation}
\begin{split}
\delta {a_m}(t)& \propto ig{\alpha _o}{e^{ - ({\gamma _m} + i{\Delta _m})t}} \cdot \\&\int\limits_0^t {ds{e^{({\gamma _m} - {\gamma _o})s}}{e^{i{\Delta _m}s}}[{e^{i(2g{\alpha _m} - {\Delta _o})s}} + {e^{ - i(2g{\alpha _m} - {\Delta _o})s}}]}
\end{split}
\end{equation}
Thus, it can be seen that for light waves, the resonance interaction occurs in ${\Delta _o} - 2g{\alpha _m} =  \pm {\Delta _m}$. The distance between its two peaks is $2{\Delta _m}$. For microwave resonance interactions occur in ${\Delta _m} =  \pm ({\Delta _o} - 2g{\alpha _m})$. The distance of its peaks is $2({\Delta _o} - 2g{\alpha _m})$. As we can know from the parameters in the above numerical analysis, optical detunning ${\Delta _o}$ is the same order of microwave detunning ${\Delta _m}$, and the driving frequency of the light wave is ${10^5}$ higher than the microwave driving frequency, so 
\begin{eqnarray}
\begin{aligned}
2{\Delta _m}/{\omega _{do}} <  < 2({\Delta _o} - 2g{\alpha _m})/{\omega _{dm}}
\end{aligned}
\end{eqnarray}
Therefore, only one peak is observed in the optical detuning curve. For the microwave detuning curve, when the driving frequency is the same as the resonant frequency of the microwave cavity, due to the strong constraint of the resonant cavity, the microwave and the light wave are difficult to interact, so the light and the microwave are not entangled. As for the microwave resonance double-peak interaction, there are other deep-seated reasons for the large difference in amplitude, which needs further study.

Further, in order to quantitatively study the quantum entanglement characteristics of light waves and microwaves, it is necessary to confirm the presence or absence of entanglement or determine the strength of the entanglement by means of actual measurement. Since it is difficult to directly measure the entanglement of light waves and microwaves, we considers the correlation matrix of two Gaussian beams measured by the scheme shown in Fig.9. As shown in Fig.9, one of the electro-optic entanglement system is coupled with another the electro-optic entanglement system by inductive coupling,and the coupling coefficient between $L_{1}$ and $L_{2}$ is adjusted so that the amplitudes of the microwave signals respectively loaded onto the electro-optic material are uniformed, and the two lasers, which are respectively the output beam of the emitted laser beam after electro-optic interaction, are detected by detectors $D_{1}$ and $D_{2}$, respectively, and then carry out correlated detection [29,30]. When the two lasers operate below the threshold, the amplitude and phase of the two laser output lights are adjusted by adjusting the transmittance and reflectivity of the beam splitter, and the correlation matrix $V$ of the bipartite quantum system composed of the two beams can be measured. Then, the matrix $A$ is obtained by the Lyapunov equation, usually a numerical solution,and the entanglement $E_{N}$ can be calculated by the formula Eq.(23).
\begin{figure}
	\centering
	\includegraphics[width=1\linewidth]{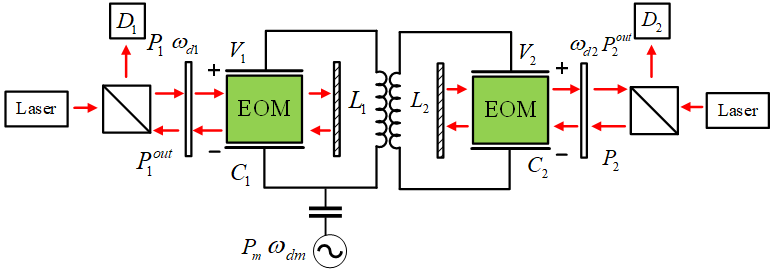}
	\caption{Optical and microwave entanglement indirect test scheme}
	\label{fig:Fig9.png}
\end{figure}

In addition, from the above numerical results that compared with the optomechanical and optoelectromechanical entanglement scheme [5-7], in order to obtain strong entanglement of the electro-optical entanglement system, the microwave and light wave input power is generally stronger. In the foregoing scheme, the general optical and microwave power are on the order of 10 mW, but in the electro-optical entanglement system, the optical and microwave power are on the order of several tens to hundreds of milliwatts, which brings some difficulties to the system to work in a low temperature environment. This problem can be solved from two aspects. One is to make the entanglement system work under the pulse duty state with lower repetition frequency to reduce the average power, and the second is consider the structure and refrigeration method of the more reasonable and efficient entanglement system.

\section{\label{sec:level1}Conclusion}
In summary, the electro-optic entanglement system proposed in this paper can realize the quantum entanglement between light waves and microwaves. The research shows that when the optical cavity and microwave cavity quality factor is large, the light wave and microwave have strong entanglement. What's more, the higher the temperature, the harder the system is to produce entanglement, but there is obvious entanglement in 15 K or even 20 K. We can see that at low temperature (i.e. close to absolute zero), working near the stable equilibrium working point, the entanglement between the light wave and the microwave is generally large, and the quantum correlation or conversion between the light wave and the microwave quantum state can be achieved, which can be used to realizing the interaction between hybrid quantum systems with different frequencies, and provide a new approach for exploiting quantum communication, quantum computing, quantum sensing and other quantum technologies in the microwave and optical bands. However, the ambient temperature discussed in this paper is still low, and its practical value is limited. Compared our work with others using the similar theoretical methods, such as, using micromirror as medium, light and light entanglement [31] logarithmic negativity is about 0.3, light field and moving cavity mirror entanglement [5] logarithmic negativity is about 0.3, atom-light-micromirror entanglement [12] logarithmic negativity is about 0.3, and the light and microwave are entangled with each other through the mechanical oscillator [7] the logarithm negativity is about 0.2. The entanglement logarithmic negativity of light and microwave entangled by the electro-optical effect is about 0.5, which may theoretically have stronger entanglement. The correctness of the theoretical analysis, the entanglement characteristics of the system and the design and preparation of the device need to be carried out in future work and experiments. In the future, we should focus on the solutions that can produce electro-optical entanglement at higher temperatures. What's more, the physical mechanism behind and related experimental verification of the characteristics of electro-optic entanglement system could be one of the future researching directions.
\section{\label{sec:level1}Appendix}
According to the Routh-Hurwitz criterion,the four conditions that need to be met for the stable operation of the electro-optic entanglement system are as follows
\begin{widetext}
	\begin{eqnarray}
	\gamma _{o}+\gamma _{m}>0
	\end{eqnarray}
	\begin{eqnarray}
	(\Delta _{m}^{2}+\gamma _{m}^{2})[(2g\alpha _{m}-\Delta _{o})^{2}+\gamma _{o}^{2}]+4g^{2}\alpha _{o}^{2}\Delta _{m}(2g\alpha _{m}-\Delta _{o})>0
	\end{eqnarray}
	\begin{eqnarray}
	\begin{aligned}
	&2(2g\alpha _{m}-\Delta _{o})^{4}\gamma _{m}+(2g\alpha _{m}-\Delta _{o})^{2}[2\Delta _{m}^{2}(\gamma _{o}+\gamma _{m})-\Delta _{m}^{2}
	+4\gamma _{o}^{2}\gamma _{m}+10\gamma _{o}\gamma _{m}^{2}+2\gamma _{m}^{3}-\gamma _{m}^{2}]\\&-4g^{2}\alpha _{o}^{2}\Delta _{m}(2g\alpha _{m}-\Delta _{o})
	+2\Delta _{m}^{4}\gamma _{o}+\Delta _{m}^{2}(2\gamma _{o}^{3}+10\gamma _{o}^{2}\gamma _{m}-\gamma _{o}^{2}+4\gamma _{o}\gamma _{m}^{2})
	+2\gamma _{o}^{4}\gamma _{m}+10\gamma _{o}^{2}\gamma _{m}^{2}(\gamma _{o}+\gamma _{m})-\\&\gamma _{o}^{2}\gamma _{m}^{2}+2\gamma _{o}\gamma _{m}^{4}>0
	\end{aligned}
	\end{eqnarray}
	\begin{eqnarray}
	\begin{aligned}
	&(2g\alpha _{m}-\Delta _{o})^{4}\gamma _{o}\gamma _{m}+2(2g\alpha _{m}-\Delta _{o})^{2}(-\Delta _{m}^{2}\gamma _{o}\gamma _{m}+\gamma _{o}^{3}\gamma _{m}+2\gamma _{o}^{2}\gamma _{m}^{2}+\gamma _{o}\gamma _{m}^{3})-4g^{2}\alpha _{o}^{2}(2g\alpha _{m}-\Delta _{o})\cdot\\&(\Delta _{m}\gamma _{o}^{2}+2\gamma _{o}\gamma _{m}+\Delta _{m}\gamma _{m}^{2})+\Delta _{m}^{4}\gamma _{o}\gamma _{m}+2\Delta _{m}^{2}(\gamma _{o}^{3}\gamma _{m}+2\gamma _{o}^{2}\gamma _{m}^{2}+\gamma _{o}\gamma _{m}^{3})\\&+\gamma _{o}^{5}\gamma _{m}+2\gamma _{o}^{4}\gamma _{m}^{2}+6\gamma _{o}^{3}\gamma _{m}^{3}+4\gamma _{o}^{2}\gamma _{m}^{4}+\gamma _{o}\gamma _{m}^{5}>0
	\end{aligned}
	\end{eqnarray}
\end{widetext}
Note that inequality (31) is trivial. When considering stability, we only need to consider the parameter regions where the other three inequalities hold.
\begin{acknowledgments}
	This work has been supported by National Key R$\& $D Program of China (2018YFA0307400); National Natural Science Foundation of China (NSFC) (61775025, 91836102). Thanks Dr. Cheng Zeng and Dayin Zhang, from the School of Optoelectronic Science and Engineering, University of Electronic Science and Technology of China, for their meaningful discussions and help.
\end{acknowledgments}

\end{document}